\documentclass[11pt,a4paper]{article}
\usepackage{jcappub}
\usepackage{amsmath,amssymb}
\usepackage{graphicx}	
\usepackage{amsfonts}

\newcommand{\dd}{\text{d}}
\newcommand{\dt}{\text{d}t}
\newcommand{\dx}{\text{d}x}
\newcommand{\overbar}[1]{
	\mkern 1.5mu\overline{\mkern-1.0mu#1\mkern-1.0mu}\mkern 1.5mu
}

\allowdisplaybreaks

\title{The double-soft limit in cosmological correlation functions and 
graviton exchange effects}

\author{Allan L. Alinea$^{a}$,}
\author{Takahiro Kubota$^{a, b, c}$}
\author{and Nobuhiko Misumi$^{a}$}

\affiliation{$^{a}$ Department of Physics, Osaka University,  Toyonaka, Osaka 560-0043, Japan}
\affiliation{$^{b}$ CELAS, Osaka University,  Toyonaka, Osaka 560-0043, Japan}
\affiliation{$^{c}$ Kavli (IPMU), The University of Tokyo, 5-1-5, Kashiwa-no-Ha,  Kashiwa City, Chiba 277-8583, Japan}

\emailAdd{kubota@celas.osaka-u.ac.jp}
\emailAdd{alinea@het.phys.sci.osaka-u.ac.jp}
\abstract
{
The graviton exchange effect on cosmological correlation functions is examined by employing  the double-soft limit technique. A new relation among correlation functions that contain the effects due to  graviton exchange diagrams 
in addition to those due to scalar-exchange and scalar-contact-interaction, is derived by  using the background field method and independently by the method of Ward identities associated with dilatation symmetry. We compare  these three terms, putting  small values for the slow-roll parameters and $(1-n_{s})\approx 0.042$,  where $n_{s}$ is the scalar spectral index. It is argued that the graviton exchange effects are more dominant than the other two and could be observed in the  trispectrum in the double-soft limit. 
Our observation strengthens the previous work by Seery, Sloth and Vernizzi, in which it has been argued that the graviton exchange dominates in the counter-collinear limit for single field slow-roll inflation.
}


\begin{document}
\maketitle

\bigskip
\section{Introduction}

The recent progress in high-precision measurement of temperature fluctuation of the cosmological microwave background radiation has enabled us to extract more and more accurate information about our Universe \cite{planck14}. The analyses of the data have been focused  mainly in terms of the scalar perturbation, but this  does not necessarily mean  that the tensor perturbation is less important. On the contrary,  thanks to the dedicated efforts by experimentalists, it is more than likely that a new development in search for $ B $-mode polarization \cite{1keckbicep2,2keckbicep2}  is just around the corner. The discovery of primordial gravitational wave responsible for the $ B $-mode polarization will be expected to open a new  era of quantum gravity, for the observation of tensor mode power spectrum will imply the existence of zero point oscillation of free gravitons. What  we would like to aim at next is a search for evidence in favor of interacting gravitons in cosmological data. 

One way to observe interacting gravitons in temperature fluctuation   is to look at \textit{trispectrum}, i.e., four-point functions of scalar perturbation \cite{cobe}-\cite{chenhuangshiuwang}. 
It has been discussed in Ref. \cite{seery2} that the four-point correlation function consists of three terms, namely, scalar-exchange, scalar-contact-interaction and the graviton exchange. Toward the computation of trispectrum Arroja and Koyama \cite{arrojakoyama} obtained the fourth order interaction Hamiltonian in the comoving gauge and the uniform curvature gauge. In the course of computation they noticed that one cannot ignore second order tensor perturbation in order to obtain the correct leading order corrections.  Seery, Sloth and Vernizzi \cite{seery2} have also  pointed out that graviton exchange effects are more dominant in  scalar four-point correlation functions  than scalar exchange. This is based on the observation that  the interaction Lagrangian between graviton and two scalars is not suppressed by any powers of slow-roll parameters while the contact interactions of three scalars are suppressed at least one power of the slow-roll parameters. 
In Refs. \cite{seeryIntermediate,seery2} the full contribution from graviton exchange to the trispectrum for a general momentum configuration was calculated. This general result was then checked in the counter-collinear limit using background field method, and it was found that in this limit graviton exchange dominates. In the present paper we will look for another way of uncovering the largeness of the graviton exchange effect in the trispectrum.

In connection with the strategy of studying graviton exchange effects on the trispectrum, let us recall that we have recently witnessed remarkable progresses in analyzing cosmological correlation functions  motivated by the consistency conditions obtained earlier in Refs. \cite{maldacena1, creminelli1}. Namely, technique of taking the soft limit \cite{cheung1}-\cite{Binosi:2015obq}  known in chiral dynamics \cite{adler}  has been applied and developed. It has been by now well-established that $(N+1)$-point correlation functions are connected with $N$-point functions by sending one of the wave vectors to zero. The connection between correlation functions may be understood either by noting \textit{Ward identities} associated with the residual conformal symmetry in three-dimensional space or by computing $N$-point functions under the influence of long-wave background fields of so-called \textit{adiabatic modes} \cite{weinberg1}. It is also possible to relate $(N+2)$-point functions to $N$-point ones by using the above method twice \cite{mirbabayi1, joyce} which is often referred to as \textit{double-soft limit technique}.

In the present paper we will extend the double-soft limit technique by including the tensor perturbation.\footnote{The basic idea has been elaborated previously by one of the authors \cite{misumi}.} We are going to derive a new relation among the trispectrum, scalar-scalar-tensor correlation, and the power spectrum, in which graviton exchange effects naturally  appear. Our new relation is derived by using two independent methods: one is the \textit{background field method} and the other is based on the \textit{Ward identities associated with the dilatation symmetry}. It is shown that the two methods lead us to the same formula as must be the case.

Our double-soft limit relation  tells us that the trispectrum consists of three types of Feynman diagrams: \textit{scalar exchange}, \textit{scalar-contact-interaction} and the \textit{graviton exchange}.  We will argue that the scalar exchange is suppressed not only by  slow-roll parameters $\epsilon$ or $\eta $, but also by  another small factor $(1- n_{s}  )$, where $n_{s}$ is the scalar spectral index \cite{pdg}
\begin{eqnarray}
	 n_{s} = 0.958 \pm 0.007\: .
	 \label{eq:numericalvalueofns}
\end{eqnarray}
The contribution due to the scalar-contact-interaction is also doubly suppressed by a factor $(1 - n_{s} )^{2}$.  The graviton exchange effect, on the other hand, is suppressed only by a single factor $\epsilon$.  This observation leads us to conclude that  we will have a good chance to observe the graviton exchange effects by looking at the double-soft limit of the trispectrum.

\bigskip
\section{Preliminaries}
In this section we review the formalism that is necessary for taking double-soft limits. Our argument can be applied to a  general class of single scalar field inflationary models, whose action together with Einstein's general relativity is given by  
\begin{eqnarray}
	S
	=
	\int \dd^{4}x \sqrt{-g} \left [
		\frac{1}{2}M_{\rm pl}^{2}R + P(\phi , X)
	\right ],
	\qquad
	X
	\equiv
	-\frac{1}{2}g^{\mu \nu}\,\nabla _{\mu}\phi \nabla_{\nu} \phi\:.
	\label{eq:kinflationmodel}
\end{eqnarray}
Here, $R$ is the scalar curvature, $M_{\rm pl}=1/\sqrt{8\pi G}$ is the Planck mass and $P(\phi , X)$ is a polynomial functional of the inflaton field $\phi$ and its canonical kinetic term $X$. This is the most general Lagrangian so far as we exclude  higher derivative terms of $\phi$.  

As for  the classical background we use the Friedmann-Lema\^itre-Robertson-Walker metric
\begin{eqnarray}
	\dd s^{2}
	=
	-\dt^{2}+a^{2}(t)\delta _{ij} \dx^{i}\dx^{j}, 
\end{eqnarray}
where the scale factor  is denoted  by $a(t)$. In order to handle quantum fluctuations around the classical background, one often uses the Arnowitt-Deser-Misner parametrization
\begin{eqnarray}
	\dd s^{2}
	=
	-N^{2}\dt^{2}+h_{ij}
	\left(\dx^{i} + N^{i} \dt\right ) 
	\left (\dx^{j} + N^{j} \dt \right ),
\end{eqnarray}
where $ N $ and $ N^i $ are the lapse and shift functions, respectively. We will use the \textit{comoving gauge} throughout the present work. The spatial part of the metric in this gauge given as 
\begin{eqnarray}
	h_{ij}
	=
	a^{2}(t)e^{2 \zeta (t, \vec{x})} 
	\left ( e^{\gamma (t, \vec{x}) }\right )_{ij}\:, 
	\label{eq:spatialmetric}
\end{eqnarray}
is parametrized by scalar ($\zeta $) and tensor ($\gamma _{ij}$) perturbations supplemented by the conditions ${\gamma ^{i}}_{i}=0$ and $\partial _{i }{ \gamma ^{i}} _{j}=0$.  

The  two-point function of $\zeta $ namely, the scalar power spectrum, has been computed in literature as 
\begin{eqnarray}
	\langle \zeta_{\vec{k}_{1}}  \zeta_{\vec{k}_{2}}  \rangle 
	= 
	(2\pi )^{3} \delta^{3}(\vec{k}_{1} + \vec{k}_{2} )
	P_{\zeta} (k_{1} ),
	\qquad
	P_{\zeta}(k)
	=
	\frac{1}{4\epsilon c_{s} k^{3} }\left ( \frac{H}{M_{\rm pl}} \right )^{2}\: , 
	\label{eq:scalarpowerspectrum}
\end{eqnarray}
where $H\equiv\dot{a}(t)/a(t)$ is the Hubble parameter and $\zeta_{\vec{k}}$ is the three-dimensional Fourier transform of $\zeta (t, \vec{x})$.  Here and hereafter the magnitudes of wave vectors $\vec{k}_{i}$ $(i=1,2, \cdots )$ are denoted by  $k_{i}=\vert \vec{k}_{i} \vert$. The sound velocity $c_{s} $ for the model of (\ref{eq:kinflationmodel}) is given by 
\begin{eqnarray}
	c_{s}^{2}
	=
	\frac{P_{, X}}{P_{, X} +2P_{, XX}}\:, 
\end{eqnarray}
where we have denoted the first and second derivatives of $P(\phi , X)$ 
with respect to  $X$ by $P_{, X}$ and $P_{, XX}$, respectively. The slow-roll parameter $\epsilon$ and $\eta$ are defined by 
\begin{eqnarray}
	\epsilon 
	\equiv
	-\frac{\dot{H}}{H^{2}}, 
	\qquad
	\eta 
	\equiv
	\frac{1}{H}\frac{\dot{\epsilon} }{\epsilon }.
	\label{eq:slowrollparameter}
\end{eqnarray}
The tensor power spectrum has also been known as 
\begin{eqnarray}
	P_{\gamma}(k) 
	=
	16 \epsilon c_{s} P_{\zeta} (k)= \frac{4}{k^{3}}\left ( \frac{H}{M_{\rm pl}} \right )^{2}\:.
	\label{eq:tensorpowerspectrum}
\end{eqnarray}
The parameters $H, \epsilon $ and $ c_s $ in (\ref{eq:scalarpowerspectrum}) and (\ref{eq:tensorpowerspectrum}) are evaluated at the time of horizon crossing.

\bigskip
\section{Background field method and graviton exchange effects} 
Now we derive the double-soft limit relations by employing the background field method. The point of this method is that correlation functions in the presence of background fields can be obtained by applying a  coordinate transformation to corresponding correlation functions in the absence of the backgrounds. This method has been fully discussed in Refs. \cite{mirbabayi1} and  \cite{joyce}, along which we will also proceed paying due attention to the tensor modes. 

Let us denote the long wavelength modes of scalar and tensor by $\zeta _{L} $ and $\gamma _{\rm L}$, respectively. The basic formula of this method for $N$-point correlation functions  at fixed time $t_{*}$ is
\begin{eqnarray}
	\langle \zeta (\vec{x}_{1}) \cdots \zeta(\vec{x}_{N}) \rangle _{\rm bg}
	=
	\langle \zeta (\vec{{\tilde x}}_{1}) \cdots \zeta(\vec{{\tilde x}}_{N}) \rangle .
	\label{eq:withwithoutrelation}
\end{eqnarray}
Here $\langle \zeta \cdots \zeta \rangle _{\rm bg }$ and $\langle \zeta \cdots \zeta \rangle $ are correlation functions with and without the background fields,  respectively. The coordinates in these correlation functions ($x^{i}$ and ${\tilde x}^{i}$) are related by 
\begin{eqnarray} 
	{\tilde x}^{i}
	&=&
	e^{\zeta _{\rm L}}{\big ( e^{ \frac{1}{2} \gamma _{\rm L}}  \big )^{i}}_{j}\,x^{j}
	\nonumber \\
	&=&
	\left(
		1 + \zeta_{\rm L} 
		+ 
		\tfrac{1}{2}{\zeta _{\rm L}}^{2} 
		+ 
		\cdots 
	\right)
	\left(
		{\delta^{i}}_{j}  
		+
		\tfrac{1}{2}{{ \gamma_{\rm L}}^{i}}_{j} 
		+ 
		\cdots 
	\right) x^{j}
	\nonumber \\
	&\equiv &
	x^{i}+\delta x^{i}.
\end{eqnarray}
Here $\zeta _{\rm L}$ and $\gamma _{\rm L}$ are the values of long-wavelength modes at some point, say, at the origin. To second order in $\zeta _{L} $ and   linear order in $\gamma _{\rm L}$, the right hand side of (\ref{eq:withwithoutrelation}) turns out to be 
\begin{eqnarray}
	\langle \zeta (\vec{{\tilde x}}_{1})  \cdots \zeta (\vec{{\tilde x}}_{N} ) \rangle 
	&=&
	\langle \zeta (\vec{x}_{1})  \cdots \zeta (\vec{x}_{N} ) \rangle
	+ 
	\sum _{a=1}^{N}\delta \vec{x}_{a}\cdot \vec{\nabla}_{a}
	\langle \zeta (\vec{x}_{1})  \cdots \zeta (\vec{x}_{N} ) \rangle
	\nonumber \\
	& & 
	+\,
	\frac{1}{2}\sum _{a, b=1}^{N}
	\delta x^{i}_{a} \delta x^{j}_{b}{\nabla _{a}}_{i} {\nabla _{b}}_{j}
	\langle \zeta (\vec{x}_{1})  \cdots \zeta (\vec{x}_{N} ) \rangle 
	+
	\cdots 
	\nonumber \\
	&=&
	\langle \zeta (\vec{x}_{1})  \cdots \zeta (\vec{x}_{N} ) \rangle
	+
	\zeta _{\rm L} \sum _{a=1}^{N} \vec{x}_{a} \cdot \vec{\nabla}_{a}
	\langle \zeta (\vec{x}_{1})  \cdots \zeta (\vec{x}_{N} ) \rangle
	\nonumber \\
	& & 
	+\, 
	\frac{\zeta ^{2}_{L}}{2}\Big (
	\sum _{a=1}^{N} 
	\vec{x}_{a} \cdot {\vec{\nabla}}_{a}
	+
	\sum _{a,b=1}^{N}
	x^{i}_{a}x^{j}_{b}{\nabla _{a}}_{i} {\nabla_{b}}_{j}
	\Big ) \langle \zeta (\vec{x}_{1})  \cdots \zeta (\vec{x}_{N} ) \rangle
	\nonumber \\
	& & 
	+\,
	\frac{1}{2}  {{ \gamma _{\rm L}}^{i}}_{j} 
	\sum _{a=1}^{N}x_{a}^{j} {\nabla _{a}}_{i}
	\langle \zeta (\vec{x}_{1})  \cdots \zeta (\vec{x}_{N} ) \rangle  
	+ 
	\cdots \: .  
\end{eqnarray}

In the Fourier-transformed space we thus have 
\begin{align}
	&\delta \langle \zeta (\vec{x}_{1}) \cdots \zeta (\vec{x}_{N}) \rangle
	\nonumber
	\\[0.5em]
	&\quad
	=
	\langle \zeta (\vec{x}_{1}) \cdots \zeta (\vec{x}_{N}) \rangle _{\rm bg}
	-
	\langle \zeta (\vec{x}_{1}) \cdots \zeta (\vec{x}_{N}) \rangle 
	\nonumber
	\\[0.5em]
	&\quad
	=
	\int \frac{d^{3}\vec{q}}{(2\pi )^{3}} 
	\int \frac{d^{3}\vec{k}_{1}}{(2\pi )^{3}}
	\cdots 
	\int \frac{d^{3}\vec{k}_{N}}{(2\pi )^{3}}\:
	(2\pi )^{3}\delta (\vec{k}_{1} + \cdots + \vec{k}_{N})
	\langle \zeta _{\vec{k}_{1}} \cdots \zeta _{\vec{k}_{N}} \rangle'
	\nonumber
	\\[0.5em]
	&\qquad	
	\times\, \bigg [
		\zeta _{\vec{q}}\sum _{a=1}^{N} \vec{k}_{a} \cdot {\vec{\nabla}}_{k_{a}}
		+
		\frac{1}{2}
		\int \frac{d^{3} \vec{Q}}{(2\pi)^{3}} \; 
		\zeta _{\vec{q}}\;\zeta _{\vec{Q}-\vec{q}}
		\Big (
			\sum _{a=1}^{N}\vec{k}_{a} \cdot \vec{\nabla }_{k_{a}}
			+
			\sum _{a, b=1}^{N}
			{k_{a}}^{i}{k_{b}}^{j}{\nabla _{k_{a}}}_{i}{\nabla _{k_{b}}}_{j} 
		\Big )
		\nonumber
		\\[0.5em]
		&\qquad\qquad
		+\,
		\frac{1}{2} {{\gamma_{\vec{q}}}^{\:i}}_{j}
		\sum _{a=1}^{N}{k_{a}}^{j} {\nabla _{k_{a}}}_{i} 
	 \bigg ]  
	 {\exp} \Big ( i \sum _{a=1}^{N}\vec{k}_{a} \cdot \vec{x}_{a}\Big )\: .
\end{align}
Here, primed correlation functions $\langle \cdots \cdots \rangle ^{\prime}$ are those whose $\vec{k}$-conserving delta function $(2\pi )^{3}\delta (\vec{k}_{1}+\cdots + \vec{k}_{N})$ has been omitted. Performing partial integration in $\vec{k}$-space, we finally get 
\begin{eqnarray}
	\delta \langle \zeta _{\vec{k}_{1} } \cdots \zeta _{\vec{k}_{N} } \rangle'
	&=&
	\lim_{\vec{q} \to 0} \bigg (
		\zeta _{\vec{q}}\,\delta _{\cal D} 
		+ 
		\frac{1}{2} \int \frac{d^{3}\vec{Q}}{(2\pi )^{3}} \:
		\zeta _{\vec{q}} \: \zeta _{\vec{Q}-\vec{q}}\: 
		\delta _{\cal D}^{2}
	\bigg ) 
	\langle \zeta _{\vec{k}_{1}} \cdots  \zeta _{\vec{k}_{N}}\rangle'
	\nonumber \\
	& &
	+\,
	\lim _{\vec{q} \to 0}  
	\frac{1}{2} {{\gamma _{\vec{q}}^{\: i}}}_{j}{\Delta ^{ j}}_{i}
	 \langle \zeta _{\vec{k}_{1}} \cdots  \zeta _{\vec{k}_{N}}\rangle'\:,
\end{eqnarray}
where our notations are
\begin{eqnarray}
	\delta_{\cal D}
	=
	-3(N-1)-\sum _{a=1}^{N}\vec{k}_{a}\cdot \vec{\nabla}_{k_{a}}, 
	\hskip0.5cm 
	{\Delta ^{j}}_{i}= - ( N - 1 ){\delta ^{j}}_{i} - \sum _{a=1}^{N} {k_{a}}^{j}\nabla _{k_{a}  i} . 
	\label{eq:deltaddeltij}
\end{eqnarray}

Multiplying both sides by two long-wavelength modes $\zeta _{\vec{q}_{1}} $ and $\zeta _{\vec{q}_{2}}$ and taking the average $\langle \cdots \cdots \rangle $, we arrive at 
\begin{eqnarray}
	\lim _{\vec{q}_{1}, \vec{q}_{2} \to 0}
	\frac{
		\langle 
			\zeta _{\vec{q}_{1}}\zeta _{\vec{q}_{2}}
			\zeta _{\vec{k}_{1}} \cdots \zeta _{\vec{k}_{N}} 
		\rangle'
	}{P_{\zeta}(q_{1})P_{\zeta}(q_{2})}
	&=&
	\frac{
		\langle 
			\zeta _{\vec{q}_{1}}\zeta _{\vec{q}_{2}}
			\zeta _{\vec{q}} 
		\rangle' 
	}{P_{\zeta }(q_{1})P_{\zeta }(q_{2})}\delta _{{\cal D}}
	\langle \zeta _{\vec{k}_{1}}\cdots \zeta _{\vec{k}_{N}} \rangle'
	+ 
	\delta ^{2}_{{\cal D}}
	\langle \zeta _{\vec{k}_{1}}\cdots \zeta _{\vec{k}_{N}} \rangle'
	\nonumber \\
	& & 
	+\, 
	\sum_{s} \frac{
		\langle 
			\zeta _{\vec{q}_{1}}\zeta _{\vec{q}_{2}}\gamma^{s} _{\vec{q}} 
		\rangle'
	}{P_{\zeta } (q_{1})P_{\zeta }(q_{2})}
	\Delta ^{s }\langle \zeta _{\vec{k}_{1}} \cdots \zeta _{\vec{k}_{N}} \rangle ^{\prime}\:, 
	\label{eq:basicformula}
\end{eqnarray}
where we have put $ {{\gamma _{ \vec{q} }}^{i}}_{j} = \sum_{s} \gamma _{\vec{q}}^{s}\:\:{{\varepsilon  }^{s i}}_{j}(\vec{q})$ and $P_{\zeta }(q_{i}) \: (\text{with\;}i=1,2\,\text{and}\;  q_{i} = |\vec{q}_{i}|) $ is the power spectrum of the scalar mode (\ref{eq:scalarpowerspectrum}). The polarization ($s= +$ or $ \times $) of the tensor mode is described by ${{\varepsilon }^{s i}}_{j}(\vec{q})$ and $\Delta ^{s}$ is a differential operator
\begin{eqnarray}
	\Delta ^{s}
	=
	-\frac{1}{2}\sum _{a=1}^{N}
	{{\varepsilon^{s}}^{i}}_{j}(\vec{q}\,)\: {k_{a}}^{j}
	{\nabla_{k_{a}}}_{i}\:.
	\label{eq:deltagamma}
\end{eqnarray}
Since the polarization tensor is traceless, the first term of ${\Delta ^{j}}_{i}$ in (\ref{eq:deltaddeltij}) does not survive in (\ref{eq:deltagamma}). The third term on the right hand side of (\ref{eq:basicformula}) is a new one in comparison  with formulae obtained in previous  literatures and represents  the graviton exchange effect. In connection with this, note  that the operation  (\ref{eq:deltagamma}) to scalar correlation functions induces a tensor-scalar cross correlation as was shown  in Refs. \cite{maldacena1, creminelli2,hinterbichler2}
\begin{eqnarray}
	\Delta ^{s} 
	\langle  \zeta _{\vec{k}_{1}}\cdots \zeta _{\vec{k}_{N}} \rangle'
	=
	- \frac{1}{2}\sum _{a=1}^{N}{{\varepsilon^{s}}^{i}}_{j} (\vec{q}\,) {k_{a}}^{j}{\nabla_{k_{a}}}_{i}
	\langle  \zeta _{\vec{k}_{1}}\cdots \zeta _{\vec{k}_{N}} \rangle'
	=
	\frac{\langle  \zeta _{\vec{k}_{1}}\cdots \zeta _{\vec{k}_{N} }\gamma _{-\vec{q}}^{s}  \rangle ^{\prime} }
	{P_{\gamma}(q)}
	+
	{\cal O}(q)\:  , 
	\label{eq:deltagammaformula}
\end{eqnarray}
where $P_{\gamma}(q)$  with $ q = |\vec{q}\,|$, is the tensor power spectrum (\ref{eq:tensorpowerspectrum}). Using (\ref{eq:deltagammaformula}), the formula (\ref{eq:basicformula}) reduces to 
\begin{eqnarray}
	\lim _{\vec{q}_{1}, \vec{q}_{2} \to 0}
	\frac{
		\langle 
			\zeta _{\vec{q}_{1}}\zeta _{\vec{q}_{2}}\zeta _{\vec{k}_{1}} \cdots \zeta _{\vec{k}_{N}} 
		\rangle'}{P_{\zeta}(q_{1})P_{\zeta}(q_{2})}
	&=&
	\frac{
		\langle 
			\zeta _{\vec{q}_{1}}\zeta _{\vec{q}_{2}}\zeta _{\vec{q}} 
		\rangle'
	}{P_{\zeta }(q_{1})P_{\zeta }(q_{2})}
	\delta _{{\cal D}}
	\langle 
		\zeta _{\vec{k}_{1}}\cdots \zeta _{\vec{k}_{N}} 
	\rangle'
	+ 
	\delta ^{2}_{{\cal D}}
	\langle \zeta _{\vec{k}_{1}}\cdots \zeta _{\vec{k}_{N}} \rangle'
	\nonumber \\
	& & 
	+\,
	\sum_{s}  
	\frac{
		\langle 
			\zeta _{\vec{q}_{1}}\zeta _{\vec{q}_{2}}\gamma^{s} _{\vec{q}} 
		\rangle' 
	}{P_{\zeta } (q_{1})P_{\zeta }(q_{2})}
	\frac{
		\langle  
			\zeta _{\vec{k}_{1}} \cdots  \zeta _{\vec{k}_{N}}\gamma ^{s}_{-\vec{q}}  
		\rangle'
	}{P_{\gamma}(q)}\:. 
\end{eqnarray}

\begin{figure}[hbt]
	\centering
	\makebox[\textwidth][c]{\includegraphics[scale=0.9]{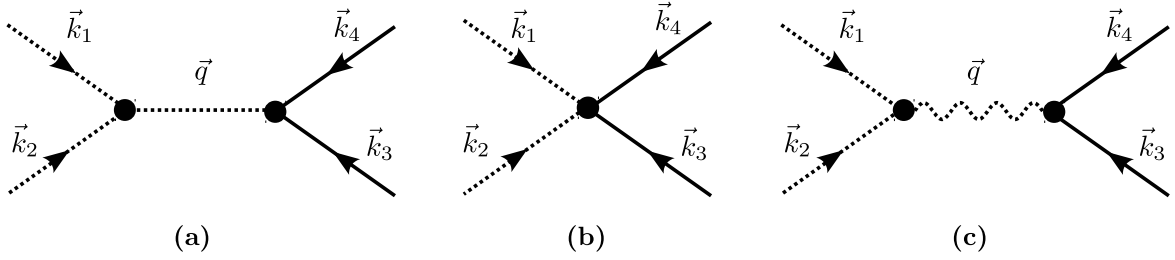}}	
	\caption{Feynman diagrams representing the contributions to the four-point function of $ \zeta _{\vec k} $ in the double-soft limit. (a) scalar exchange, (b) contact interaction and (c) graviton exchange diagrams contributing to the trispectrum. Dashed and solid lines denote soft and hard scalars, respectively. The wavy line in (c) corresponds to a  graviton exchange. }
	\label{fig:1}
	\bigskip
\end{figure}

For the trispectrum we get a useful formula (replacing soft wave vectors $\vec{q}_{1}$ and $\vec{q}_{2}$ by $\vec{k}_{1}$ and $\vec{k}_{2}$, respectively,)
\begin{eqnarray}
	\lim _{\vec{k}_{1}, \vec{k}_{2} \to 0}
	\frac{
		\langle 
			\zeta _{\vec{k}_{1}}\zeta _{\vec{k}_{2}}
			\zeta _{\vec{k}_{3}}  \zeta _{\vec{k}_{4}} 
		\rangle'
	}{P_{\zeta}(k_{1})P_{\zeta}(k_{2})}
	&=&
	\frac{
		\langle 
			\zeta _{\vec{k}_{1}}\zeta _{\vec{k}_{2}}\zeta _{\vec{q}} 
		\rangle' 
	}{P_{\zeta }(k_{1})P_{\zeta }(k_{2})}
	\delta _{{\cal D}} P_{\zeta}(k_{3})
	+
	\delta ^{2}_{{\cal D}} P_{\zeta}(k_{3})
	\nonumber \\
	& & 
	+\,
	\sum _{s}  
	\frac{
		\langle 
			\zeta _{\vec{k}_{1}}\zeta _{\vec{k}_{2}}\gamma^{s} _{\vec{q}} 
		\rangle'
	}{P_{\zeta } (k_{1})P_{\zeta }(k_{2})}
	\frac{
		\langle 
			\zeta _{\vec{k}_{3}}   \zeta _{\vec{k}_{4}}	\gamma ^{s}_{-\vec{q}} 
		\rangle'}{P_{\gamma}(q)}\:. 
	\label{eq:basicformula1}
\end{eqnarray}
The first term on the right hand side of  (\ref{eq:basicformula1})  comes from  the \textit{scalar exchange} (\textbf{Fig}. \ref{fig:1}(a))  and the second term is due to the \textit{scalar-contact-interaction} (\textbf{Fig.} \ref{fig:1}(b)). These two terms have been discussed in previous literatures \cite{mirbabayi1, joyce}, while the third term is a new addition in the present work and corresponds to \textit{graviton exchange effect} (\textbf{Fig.} \ref{fig:1}(c)). Note that, in the case of single-soft limit, soft internal graviton lines have been studied briefly in Refs. \cite{creminelli2, hinterbichler2}.

\bigskip
\section{Ward identities and graviton exchange effects}
In this section we would like to derive the formula  (\ref{eq:basicformula1}) by using the Ward identities associated with dilatation symmetry. The theoretical framework for this purpose has been developed in  Refs. \cite{goldberger, berezhiani}, the starting point of which is the one-particle irreducible part $\Gamma [\zeta , \gamma ]$, i.e., a generating functional in three dimensional field theory at fixed time $t_{*}$:
\begin{eqnarray}
	Z[J, T^{ij} ]
	&=&
	\int {\cal D}\zeta (t_{*}, \vec{x}) {\cal D} \gamma _{ij} (t_{*}, \vec{x} )
	P[t_{*}, \zeta , \gamma_{ij}]\:
	\nonumber \\
	& & 
	\times\, 
	\exp\left \{
		\int d^{3}\vec{x}\, \big[
			\zeta (t_{*}, \vec{x} ) J(\vec{x}) 
			+ 
			\gamma_{ij}(t_{*}, \vec{x} )T^{ij}(\vec{x} ) 
		\big]
	\right \}\: .
\end{eqnarray}
Here the probability measure $P[t_{*}, \zeta , \gamma_{ij}]$ is given by an integral over the vacuum wave functional at time $t_{*}$, and the formulation in Ref. \cite{goldberger} has been slightly extended to include the tensor mode path integral. The generating functional of one-particle irreducible  Green's functions is given as usual by a Legendre transformation
\begin{eqnarray}
	\Gamma [\overbar{\zeta} , \overbar{\gamma} _{ij}  ] 	
	=
	\ln Z[J, T^{ij}]
	-
	\int d^{3}\vec{x} \left \{  
		J(\vec{x} ) \overbar{\zeta} (t_{*}, \vec{x} )
		+ 
		T^{ij} (\vec{x}) \overbar{\gamma} _{ij} (t_{*}, \vec{x}) 
	\right \}\:, 
\end{eqnarray}
where we have defined
\begin{eqnarray}
	\overbar{\zeta} (t_{*}, \vec{x})
	=
	\frac{\delta {\rm ln} Z[J, T^{ij} ]}{\delta J(\vec{x})}, 
	\qquad
	\overbar{\gamma}_{ij}(t_{*}, \vec{x} )
	=
	\frac{\delta {\rm ln} Z[J, T^{ij} ]}{\delta T^{ij} (\vec{x})}\:. 
\end{eqnarray}

The scalar $N$-point Green's function $\Gamma ^{(N)}$ is defined by 
\begin{eqnarray}
	\frac{
		\delta ^{N} \Gamma [ \overbar{\zeta} , \overbar{\gamma}_{ij}  ]
	}{
		\delta { \overbar{\zeta} _{\vec{k}_{1}}} 
		\cdots 
		{\delta \overbar{\zeta }_{\vec{k}_{N}}} 
	}
	=
	(2\pi )^{3}\delta ^{3}(\vec{k}_{1}+ \cdots  + \vec{k}_{N})
	\Gamma ^{(N)}( \vec{k}_{1}, \cdots , \vec{k}_{N})\:, 
\end{eqnarray}
and it has been shown that the dilatation symmetry\footnote{For an alternative approach of deriving the four-point function using Ward identities involving special conformal transformation, see Ref. \cite{Ghosh:2014kba}.} controls the relation between $N$- and $(N+1)$-point functions \cite{goldberger}
\begin{eqnarray}
	\lim _{\vec{q} \to 0}\:\Gamma ^{(N+1)} 
	(\vec{q}, \vec{k}_{1}, \cdots , \vec{k}_{N})
	=
	\left ( 3 - {\cal D}_{N} \right )
	\Gamma ^{(N)}(  \vec{k}_{1}, \cdots , \vec{k}_{N})\:, 
	\label{eq:singlesoftlimitrelation}
\end{eqnarray}
where $ \mathcal D_N $ is the dilatation derivative defined as
\begin{eqnarray}
	{\cal D}_{N}
	\equiv
	\sum _{a=1}^{N}\vec{k}_{a} \cdot \vec{\nabla}_{k_{a}} .
\end{eqnarray}
Although the tensor perturbation was not the focus of attention in previous literatures, the formula  (\ref{eq:singlesoftlimitrelation}) is still valid in this form even after we consider tensor contributions. For $N=2$ case, let us recall the relation
\begin{eqnarray}
	\Gamma ^{(2)} (k) 
	=
	-\frac{1}{P_{\zeta}(k)}\:, 
	\label{eq:gammaandp}
\end{eqnarray}
and note that  (\ref{eq:singlesoftlimitrelation}) is reduced to the celebrated Maldacena's relation \cite{maldacena1}
\begin{eqnarray}
	\lim_{\vec{q} \to 0} 
	\Gamma^{(3)}( \vec{q}, \vec{k}_{1}, \vec{k}_{2} )
	=
	(3-{\cal D}_{2}) 
	\Gamma ^{(2)} (\vec{k}_{1}, \vec{k}_{2} )
	=
	-\frac{1}{P_{\zeta} (k)^{2}} 
	\left ( 3 + {\cal D}_{1} \right )
	P_{\zeta}(k)\: , 
	\label{eq:maldacenarelation}
\end{eqnarray}
where $k=\vert \vec{k}_{1} \vert = \vert \vec{k}_{2} \vert $. 
 
Suppose that we use the relation (\ref{eq:singlesoftlimitrelation}) twice, then we have
\begin{eqnarray}
	\lim _{\vec{q}_{1}, \vec{q}_{2}  \to 0}\:
	\Gamma ^{(N+2)} (
		\vec{q}_{1}, \vec{q}_{2}, \vec{k}_{1}, \cdots , \vec{k}_{N}
	)
	=
	\left ( 3 - {\cal D}_{N}\right )
	\left ( 3 - {\cal D}_{N} \right )
	\Gamma ^{(N)}(  \vec{k}_{1},  \cdots , \vec{k}_{N})\:. 
	\label{eq:doublesoftlimitgeneral}
\end{eqnarray}
We now make use of  (\ref{eq:singlesoftlimitrelation}) and (\ref{eq:doublesoftlimitgeneral}) in order to disentangle the four-point correlation function decomposed into 1PI Green's functions 
\begin{eqnarray}
	& & 
	\hskip-1.5cm
	\langle 
		\zeta _{\vec{k}_{1}} \zeta _{\vec{k}_{2}} 
		\zeta _{\vec{k}_{3}} \zeta _{\vec{k}_{4}} 
	\rangle  
	=
	(2\pi )^{3}\delta ^{3}(
		\vec{k}_{1} + \vec{k}_{2}  + \vec{k}_{3} + \vec{k}_{4} 
	)
	P_{\zeta } (k_{1})P_{\zeta } (k_{2})P_{\zeta } (k_{3})P_{\zeta } (k_{4})
	\nonumber \\
	& \times &
	\bigg [\,
		\Gamma ^{(4)} (\vec{k}_{1}, \vec{k}_{2}, \vec{k}_{3}, \vec{k}_{4}) 
		\nonumber \\
		& & 
		+\,
		\Gamma ^{(3)}(\vec{k}_{1}, \vec{k}_{2}, \vec{q}\, )
		P_{\zeta} (q)
		\Gamma ^{(3)} (\vec{k}_{3}, \vec{k}_{4}, -\vec{q}\, )
		\nonumber \\[0.5em]
		& &
		+\,
		\Gamma ^{(3)}(\vec{k}_{1}, \vec{k}_{3}, -\vec{k}_{1}-\vec{k}_{3}  ) 
		P_{\zeta} (|\vec{k}_{1}+\vec{k}_{3}|)
		\Gamma ^{(3)} (\vec{k}_{2}, \vec{k}_{4}, \vec{k}_{1}+\vec{k}_{3} )
		\nonumber \\[0.5em]
		& & 
		+\,
		\Gamma ^{(3)}(\vec{k}_{1}, \vec{k}_{4}, -\vec{k}_{1}-\vec{k}_{4}  )
		P_{\zeta} (|\vec{k}_{1}+\vec{k}_{4}|)
		\Gamma ^{(3)} (\vec{k}_{2}, \vec{k}_{3}, \vec{k}_{1}+\vec{k}_{4} )
		\nonumber \\[0.5em]
		& &
		+\,
		\sum_{s} 
		\Gamma ^{(3)}_{\zeta \zeta \gamma}(\vec{k}_{1}, \vec{k}_{2};\vec{q},s) P_{\gamma}(q)
		\Gamma ^{(3)}_{\zeta \zeta \gamma}(  \vec{k}_{3}, \vec{k}_{4};-\vec{q},s)
		\nonumber \\
		& &
		+\,
		\sum_{s}  
		\Gamma ^{(3)}_{\zeta \zeta \gamma}
		(\vec{k}_{1}, \vec{k}_{3};  -\vec{k}_{1} - \vec{k}_{3}, s ) 
		P_{\gamma}(|\vec{k}_{1} + \vec{k}_{3}|)
		\Gamma ^{(3)}_{\zeta \zeta \gamma}
		( \vec{k}_{2}, \vec{k}_{4};     \vec{k}_{1}+\vec{k}_{3}, s )
		\nonumber \\
		& &
		+\,
		\sum _{s}  
		\Gamma ^{(3)}_{\zeta \zeta \gamma}
		(\vec{k}_{1}, \vec{k}_{4};  -\vec{k}_{1}-\vec{k}_{4}, s ) 
		P_{\gamma} (\vert \vec{k}_{1}+\vec{k}_{4} \vert ) 
		\Gamma ^{(3)}_{\zeta \zeta \gamma}
		(\vec{k}_{2}, \vec{k}_{3};  \vec{k}_{1}+\vec{k}_{4}, s )
	\bigg ] , 
	\label{eq:4zetacorrelation}
\end{eqnarray}
where $\vec{q}=-\vec{k}_{1}-\vec{k}_{2}=\vec{k}_{3}+\vec{k}_{4}$. The last three terms on the right hand side  of (\ref{eq:4zetacorrelation}) are the graviton exchange effect. The connection between 1-graviton and 2-scalar  1PI part $\Gamma ^{(3)}_{\zeta \zeta \gamma}$ and correlation function is given by
\begin{eqnarray}
	\langle 
		\zeta _{\vec{k}_{1}}\zeta _{\vec{k}_{2}}  \gamma ^{s}_{\vec{q}}  
	\rangle'
	=
	P_{\zeta }(k_{1})P_{\zeta }(k_{2})P_{\gamma }(q)
	\Gamma ^{(3)}_{ \zeta \zeta \gamma }
	(\vec{k}_{1}, \vec{k}_{2};  \vec{q}, s)\: .
	\label{eq:gammazetazetagamma}
\end{eqnarray}

In the double-soft limit ($\vec{k}_{1} \to 0$, $\vec{k}_{2} \to 0$), $\vec{q} $ also becomes soft and among the three  terms of graviton exchange the one containing $P_{\gamma} (q)$ becomes the most dominant, i.e., 
\begin{eqnarray}
	\lim _{\vec{k}_{1}, \vec{k}_{2} \to 0} 
	\frac{
		\langle 
			\zeta _{\vec{k}_{1}} \zeta _{\vec{k}_{2}} 
			\zeta _{\vec{k}_{3}} \zeta _{\vec{k}_{4}} 
		\rangle'
	}{P_{\zeta}(k_{1}) P_{\zeta} (k_{2})}
	&\approx &
	P_{\zeta } (k_{3})P_{\zeta } (k_{4})
	\bigg [
		(3 - {\cal D}_{2})(3-{\cal D}_{2})
		\Gamma ^{(2)}( \vec{k}_{3}, \vec{k}_{4} )
		\nonumber \\
		& &
		\qquad
		+\, 
		\frac{
			\langle 
				\zeta_{\vec{k}_{1}}  \zeta_{\vec{k}_{2}} \zeta_{\vec{q}} 
			\rangle'
		}{P_{\zeta}(k_{1}) P_{\zeta} (k_{2}) } 
		(3-{\cal D}_{2}) \Gamma^{(2)}(\vec{k}_{3}, \vec{k}_{4} )
		\nonumber \\
		& &
		\qquad
		+\,
		2\left ( 3-{\cal D}_{2} \right )
		\Gamma ^{(2)} (\vec{k}_{3}, \vec{k}_{4} )P_{\zeta} (k_{3})
		\left ( 3-{\cal D}_{2} \right )
		\Gamma ^{(2)} (\vec{k}_{3}, \vec{k}_{4} )
		\nonumber \\
		& & 
		\qquad
		+\,
		\sum_{s} 
		\Gamma ^{(3)}_{\zeta \zeta \gamma}
		(\vec{k}_{1}, \vec{k}_{2};  \vec{q}, s ) 
		P_{\gamma}(q)
		\Gamma ^{(3)}_{\zeta \zeta \gamma}
		(\vec{k}_{3}, \vec{k}_{4};  -\vec{q}, s )
	\bigg ] \:. 
	\label{eq:4zetacorrelation2}
\end{eqnarray}
Using   (\ref{eq:gammaandp})  and (\ref{eq:gammazetazetagamma}) in (\ref{eq:4zetacorrelation2}), we immediately arrive at 
\begin{eqnarray}
	\lim _{\vec{k}_{1}, \vec{k}_{2} \to 0}
	\frac{
		\langle 
			\zeta _{\vec{k}_{1}}\zeta _{\vec{k}_{2}}
			\zeta _{\vec{k}_{3}} \zeta _{\vec{k}_{4}} 
		\rangle'
	}{P_{\zeta }(k_{1})P_{\zeta }(k_{2})}
	&=&
	-
	\frac{
		\langle 
			\zeta _{\vec{k}_{1}}\zeta _{\vec{k}_{2}}\zeta _{\vec{q}} 
		\rangle'
	}{P_{\zeta }(k_{1})P_{\zeta }(k_{2})}
	(3+{\cal D}_{1})
	P_{\zeta }(k_{3})
	+ 
	\left(9+6{\cal D}_{1} + {\cal D}_{1}^{2} \right )
	P_{\zeta }(k_{3})
	\nonumber \\
	& & 
	+\,
	\sum _{s}  
	\frac{
		\langle 
			\zeta _{\vec{k}_{1}}\zeta _{\vec{k}_{2}} \gamma^{s} _{\vec{q}} 
		\rangle'
	}{P_{\zeta }(k_{1})P_{\zeta }(k_{2})}
	\frac{
		\langle  
			\zeta _{\vec{k}_{3}} \zeta _{\vec{k}_{4}} \gamma^{s}_{-\vec{q}}
		\rangle'
	}{P_{\gamma} (q)}\:,
	\label{eq:basicformula2}
\end{eqnarray}
which we find identical  with (\ref{eq:basicformula1}) if we notice the relations
\begin{eqnarray}
	\delta _{\cal D}
	\langle \zeta_{\vec{k}_{3}} \zeta_{\vec{k}_{4}} \rangle ^{\prime}
	=
	-\left ( 3+ {\cal D}_{1} \right ) P_{\zeta}(k_{3}), 
	\qquad
	\delta _{\cal D}^{2} 
	\langle \zeta_{\vec{k}_{3}} \zeta_{\vec{k}_{4}} \rangle ^{\prime}
	=
	 \left (9 + 6{\cal D}_{1} +{\cal D}_{1}^{2} \right ) 
	 P_{\zeta} (k_{3})\: .
\end{eqnarray}

\bigskip
\section{Graviton exchange diagrams in the double-soft limit}
We have seen that the graviton exchange effect in the scalar trispectrum naturally appears in the double-soft limit relation as the third term  in   (\ref{eq:basicformula1}) and in (\ref{eq:basicformula2}).  For an illustrative   purpose, let us confirm these contributions by making an explicit use of the formulae available in the literatures. Seery et al. \cite{seery2} computed the trispectrum for  single field inflation with the canonical kinetic term, i.e., $P(X, \phi) =X-V(\phi)$ with $c_{s}^2=1$. They used the so-called \textit{flat gauge} $h_{ij} =a^{2}(t) \left (  e^{\gamma}\right )_{ij}$, in which the inflation field $\phi $ is split into the classical part $\phi _{c}$ and the quantum fluctuation part $\varphi $, i.e.,
\begin{eqnarray}
	\phi (\tau, \vec{x})
	=
	\phi_{c}(\tau ) + \varphi (\tau, \vec{x})\:. 
\end{eqnarray}
Here and hereafter we will use the conformal time $\tau$ defined as $\dd\tau \equiv \dt/a(t)$, instead of the coordinate time $t$. The correlation functions $\langle \zeta \cdots \zeta \rangle $ are obtained from those of $\varphi$  by a substitution  
\begin{eqnarray}
	\varphi \to \sqrt{2\epsilon}\, M_{\rm pl} \: \zeta , 
	\label{eq:substitution}
\end{eqnarray}
where $\epsilon $ is the slow-roll parameter in (\ref{eq:slowrollparameter}).

According to their application of the \textit{in-in} or \textit{Schwinger-Keldysh formalism}, the graviton exchange (GE) contribution to four-point correlator of $\varphi ^{,}s$ due to the interaction Hamiltonian
\begin{eqnarray}
	H_{\rm int}
	=
	-\frac{1}{2} \int d^{3}\vec{x} \: 
	a^{2} \gamma ^{ij} 
	\partial _{i}\varphi\,
	\partial_{j}\varphi,
\end{eqnarray}
is expressed as 
\begin{eqnarray}
	& & \hskip-1.5cm 
	\langle 
		\varphi_{\vec{k}_{1}}  \varphi_{\vec{k}_{2}}  
		\varphi_{\vec{k}_{3}}  \varphi_{\vec{k}_{3}} 
	\rangle  _{\rm GE}
	\nonumber \\
	&=&
	(2 \pi)^{3} 
	\delta^{3}( \vec{k}_{1}+ \vec{k}_{2}+\vec{k}_{3} +\vec{k}_{4} ) 
	\sum _{s}\varepsilon_{ij}^{s}(\vec{q}\,) 
	(\vec{k}_{1})^{i} (\vec{k}_{2})^{j}
	\varepsilon_{lm}^{s}(-\vec{q}\,) 
	(\vec{k}_{3})^{l} 
	(\vec{k}_{4})^{m}
	\nonumber \\
	& &
	\times\,  
	\frac{1}{(H_{*})^{4}} 
	\int_{-\infty}^{\tau_{*}} 
	\frac{d\tau}{\tau^{2}} 
	\int_{-\infty}^{\tau} 
	\frac{d\tau'}{{\tau'}^2}
	\: 
	{\rm Im} [
		G^{>}_{k_{1}}(\tau _{*}, \tau ) 
		G^{>}_{k_{2}}(\tau _{*}, \tau )
	]
	\nonumber \\
	& &
	\hskip4cm 
	\times\,
	{\rm Im} [
		G^{>}_{k_{3}}(\tau _{*}, \tau^{\prime} ) 
		G^{>}_{k_{4}}(\tau _{*}, \tau^{\prime} ) 
		F_{q}^{s >}(\tau , \tau^{\prime} )
	]
	\nonumber \\
	& &
	+\,
	({\rm permutations} )\:.
	\label{eq:seery}
\end{eqnarray}
Note that $\vec{q}= - \vec{k}_{1} - \vec{k}_{2} =  \vec{k}_{3} + \vec{k}_{4}$  is the momentum carried by the exchanged graviton with polarization tensor $\varepsilon ^{s}_{ij}(\vec{q}\, )$. The Fourier transform of $\varphi (\tau,\vec{x})$ is denoted by $\varphi_{\vec{k}}(\tau )$. The scalar and tensor two-point functions of the ``$> $" type, i.e., 
\begin{eqnarray}
	G^{>}_{k}(\tau , \tau')
	=
	U_{k}(\tau) U_{k}^{*}(\tau'), 
	\qquad
	F_{k}^{s >} (\tau , \tau') 
	= 
	V_{k}^{s} (\tau ) V_{k}^{s *} (\tau'),
\end{eqnarray}
are defined in terms of the solutions to the equations of motion
\begin{eqnarray}
	U_{k}(\tau ) 
	= 
	\frac{H_{*}}{\sqrt{2k^{3}}}(1 - ik \tau ) e^{ik\tau }, 
	\qquad
	V_{k}^{s} (\tau )
	= 
	\frac{1}{ \sqrt{k^{3}}} 
	\left ( \frac{H_{*} }{M_{\rm pl} } \right )
	(1-ik\tau )e^{ik\tau}\: . 
	\label{eq:mode}
\end{eqnarray}
The Hubble parameter at the time of horizon crossing ($\tau = \tau_{*}$) is denoted by  $H_{*}$. Recall that the functions (\ref{eq:mode}) appear in the mode expansion of $\varphi $ and $\gamma_{ij}$, respectively:
\begin{eqnarray}
	\varphi (\tau, \vec{x})
	&=& 
	\int \frac{d^{3}\vec{k}}{(2\pi )^{3}}
	\left [
		a_{\vec{k}} U_{k}(\tau) 
		+ 
		a_{-\vec{k}}^{\dag} U^{*}_{k}(\tau)
	\right ] e^{i\vec{k}\cdot \vec{x}}\:, 
	\\
	\gamma_{ij} (\tau , \vec{x})
	&=& 
	\sum_{s=+, \times } 
	\int \frac{d^{3}\vec{k}}{(2\pi )^{3}}
	\left [
		b^{s}_{\vec{k}} \: 
		\varepsilon_{ij}^{s}(\vec{k})V_{k}(\tau) 
		+ 
		b_{-\vec{k}}^{s \dag} \: 
		\varepsilon_{ij}^{s}(-\vec{k})V^{*}_{k}(\tau)
	\right ] e^{i\vec{k}\cdot \vec{x}}\:. 
\end{eqnarray}
Here  $a_{\vec{k}}$ ($a_{\vec{k}}^{\dag}$) and $b_{\vec{k}}^{s}$ ($b_{\vec{k}}^{s \dag }$) are annihilation (creation) operators of inflaton and graviton with polarization $s$, respectively

In the following we would like to examine the structure of (\ref{eq:seery}) in the soft limit $\vec{k}_{1}, \vec{k}_{2} \to 0$ and therefore the soft  exchanged graviton limit, i.e.,   $\vec{q}  \to 0$. First of all we note that the tensor two-point function is approximated by 
\begin{eqnarray}
	F_{q}^{s >}(\tau , \tau ^{\prime}) 
	\approx  
	\frac{1}{q^{3}} 
	\left ( \frac{H_{*} }{M_{\rm pl} } \right )^{2}
	=
	\frac{1}{4} P_{\gamma }(q)\: .
\end{eqnarray}
Under this approximation we note that the integrand in (\ref{eq:seery}) is symmetric under the simultaneous exchange $(\vec{k}_{1}, \vec{k}_{2} ) \leftrightarrow (\vec{k}_{3}, \vec{k}_{4} )$ and $\tau \leftrightarrow \tau^{\prime}$. Then we are allowed to extend the integration region of $\tau^{\prime} $ to $( -\infty , \tau_{*} ]$ by considering a pair of two terms in (\ref{eq:seery}) simultaneously. As one may have expected, the integration in (\ref{eq:seery}) becomes  essentially  the same as the one  discussed in Ref. \cite{maldacena1}, namely 
\begin{align}
	\int ^{\tau_{*}}_{-\infty} \frac{\dd\tau }{\tau ^{2}} \:\: 
	{\rm Im} [
		G^{>}_{k_{1}}(\tau _{*}, \tau ) 
		G^{>}_{k_{2}}(\tau _{*}, \tau )
	]
	&=
	\frac{H_{*}^{2}}{2k_{1}^{3}} 
	\frac{H_{*}^{2}}{2k_{2}^{3}}
	\int ^{\tau_{*}}_{-\infty} \frac{\dd\tau }{\tau ^{2}}  \:\:
	{\rm Im}\big[
		(1-ik_{1} \tau_{*})(1+ik_{1}\tau) 
		\nonumber 
		\\
		&\quad 
		\times\, 
		(1-ik_{2}\tau_{*})(1+ik_{2}\tau  )
		e^{i(k_{1}+k_{2})(\tau_{*} - \tau )}
	\big]
	\nonumber 
	\\[0.5em]
	&=
	\frac{H_{*}^{2}}{2k_{1}^{3}} 
	\frac{H_{*}^{2}}{2k_{2}^{3}}
	\left (
		k_{1}+k_{2} -\frac{k_{1}k_{2}}{k_{1}+k_{2}}
	\right )
	\nonumber \\[0.5em]
	\int ^{\tau_{*}}_{-\infty} \frac{\dd\tau }{\tau ^{2}} \:\: 
	{\rm Im} [
		G^{>}_{k_{1}}(\tau _{*}, \tau ) 
		G^{>}_{k_{2}}(\tau _{*}, \tau )
	]\bigg|_{k_1 = k_2}
	&=
	\frac{3H_{*}^{4}}{8}  \frac{1}{k_{1}^{5}}, 
	\label{eq:maldacenaintegrationformula}
\end{align}
where we have set $k_{1}=k_{2} $ at the last equality. 

On employing the formula (\ref{eq:maldacenaintegrationformula}),  the double integration in (\ref{eq:seery}) turns out to be 
\begin{eqnarray}
	& & 
	\hskip-1.5cm
	\sum _{s}
	\varepsilon_{ij}^{s}(\vec{q}\,) (\vec{k}_{1})^{i} (\vec{k}_{2})^{j}
	\varepsilon_{lm}^{s}(-\vec{q}\,) (\vec{k}_{3})^{l} (\vec{k}_{4})^{m}
	\frac{1}{(H_{*})^{4}} 
	\int_{-\infty}^{\tau_{*}} \frac{\dd\tau }{\tau^{2}} 
	\int_{-\infty}^{\tau_{*}} 
	\frac{\dd\tau'}{{\tau'}^2}
	\nonumber \\
	& & 
	\: \times\, 
	{\rm Im} \big[
		G^{>}_{k_{1}}(\tau _{*}, \tau ) 
		G^{>}_{k_{2}}(\tau _{*}, \tau )
	\big]
	{\rm Im} \big[
		G^{>}_{k_{3}}(\tau _{*}, \tau^{\prime} ) 
		G^{>}_{k_{4}}(\tau _{*}, \tau^{\prime} ) 
		F_{q}^{s >}(\tau , \tau^{\prime} )
	\big]
	\nonumber \\[0.5em]
	& \approx &
	\frac{1}{4}P_{\gamma}(q)(H_{*} )^{4}
	\sum _{s}
	\varepsilon_{ij}^{s}(\vec{q}\,) (\vec{k}_{1})^{i} (\vec{k}_{2})^{j}
	\varepsilon_{lm}^{s}(-\vec{q}\,) (\vec{k}_{3})^{l} (\vec{k}_{4})^{m}
	\left (  
		\frac{3}{8 k_{1}^{5} } \right ) 
		\left (  \frac{3}{ 8 k_{3}^{5} } \right ) 
	\nonumber \\
	&=&
	\frac{1}{256} P_{\gamma}(q)  (H_{*})^{4}
	\sum _{s} 
	\left [ 
		\varepsilon_{ij}^{s}(\vec{q}\,) {k_{1}}^{i} 
		{ \nabla_{\vec{k}_{1}}}^{j} 
		\frac{1}{k_{1}^{3}} 
	\right ]
	\left [
		\varepsilon_{lm}^{s}(-\vec{q}\,) {k_{3}}^{l} 
		{ \nabla_{\vec{k}_{3}}}^{m} 
		\frac{1}{k_{3}^{3}} 
	\right ]
	\nonumber \\
	&=&
	\frac{1}{16} \epsilon^{2} M_{\rm pl}^{4} \: P_{\gamma}(q) 
	\sum _{s} 
	\left [
		\varepsilon_{ij}^{s}(\vec{q}\,) {k_{1}}^{i} 
		{ \nabla_{\vec{k}_{1}}}^{j} P_{\zeta}(k_{1})  
	\right ]
	\left [
		\varepsilon_{lm}^{s}(-\vec{q}\,) {k_{3}}^{l} 
		{ \nabla_{\vec{k}_{3}}}^{m} P_{\zeta}(k_{3})  
	\right ]
	\nonumber \\
	&=&
	\frac{1}{4}  \epsilon^{2} M_{\rm pl}^{4} \: P_{\gamma}(q) 
	\sum _{s} 
	\frac{
		\langle 
			\zeta_{\vec{k}_{1}} \zeta _{\vec{k}_{2}} \gamma _{\vec{q}}^{s} 
		\rangle'
	}{P_{\gamma}(q)}
	\frac{ 
		\langle   
			\zeta_{\vec{k}_{3}} \zeta _{\vec{k}_{4}}\gamma _{-\vec{q}}^{s}  
		\rangle'}
	{P_{\gamma}(q)}\:, 
\end{eqnarray}
where use has been made of the relation (\ref{eq:deltagammaformula}) with $N=2$. After  employing the replacement (\ref{eq:substitution}) and taking into account  the combinatoric factors we end up with the formula\footnote{Note that (\ref{GEEquation}) is consistent with the formula for graviton-exchange contribution derived in Ref. \cite{seery2} using the background field method.}
\begin{eqnarray}
	\label{GEEquation}
	\langle 
		\zeta _{\vec{k}_{1} }\zeta _{\vec{k}_{2}} 
		\zeta _{\vec{k}_{3}} \zeta _{\vec{k}_{4}} 
	\rangle ^{\prime} _{\rm GE}
	=
	\sum_{s}
		\langle 
			\zeta _{\vec{k}_{1}} \zeta _{\vec{k}_{2}} \gamma _{\vec{q} }^{s} 
		\rangle'
	\frac{1}{P_{\gamma}(q)} 
	\langle  
		\zeta_{\vec{k}_{3} } \zeta _{\vec{k}_{4}} \gamma _{-\vec{q}}^{s}    
	\rangle'\:,
\end{eqnarray}
and this confirms  the third term  in   (\ref{eq:basicformula1}) and in  (\ref{eq:basicformula2}) via explicit computation.

\bigskip
\section{Discussion}
Let us now discuss the relative magnitudes of the three terms on the right hand side of (\ref{eq:basicformula2}). Just for simplicity, we will work for the case of  the single scalar standard  inflation  with $P(X, \phi ) = X -V(\phi) $ (corresponding to $c_{s}=1$) and the slow-roll approximation will be used. First of all, note  the relations
\begin{eqnarray}
	(3+{\cal D}_{1})P_{\zeta}(k)
	=
	(n_{s} -1) P_{\zeta}(k), 
	\qquad 
	(9+ 6{\cal D}_{1} +{\cal D}_{1}^{2}) P_{\zeta}(k)
	=
	(n_{s}-1)^{2}P_{\zeta}(k)\:, 
\end{eqnarray}
where $n_{s}$ is given by (\ref{eq:numericalvalueofns}). The second term in (\ref{eq:basicformula2}) is therefore estimated as 
\begin{eqnarray}
	(9+6{\cal D}_{1}+{\cal D}_{1}^{2})P_{\zeta}(k) 
	=
	( 1.76 \times 10^{-3} )\times 
	\left ( \frac{1- n_{s}}{0.042} \right )^{2} 
	P_{\zeta}(k)
\end{eqnarray}

The scalar three-point functions have been computed as \cite{maldacena1, seery1, seeryIntermediate, chenhuangshiuwang}
\begin{eqnarray}
	\langle 
		\zeta _{\vec{k}_{1}} \zeta _{\vec{k}_{2}} \zeta _{\vec{k}_{3}} 
	\rangle'
	&=&
	4P_{\zeta}(k_{1}P_{\zeta}(k_{2})\frac{{\cal A}_{\zeta} }{k_{3}^{3}}
	\\
	{\cal A}_{\zeta}
	&=&
	\epsilon \bigg (
		\hspace{-0.5em}
		-\frac{1}{8} \sum_{a=1}^{3} k_{a}^{3} 
		+
		\frac{1}{8}\sum_{a\neq b} k_{a}k_{b}^{2} 
		+ 
		\frac{1}{K}\sum _{a>b}k_{a}^{2}k_{b}^{2}
	\bigg )
	+
	\frac{\eta}{8} \sum_{a=1}^{3}k_{a}^{3}\:,
	\label{bispectrum1}
\end{eqnarray}
where $K=k_{1}+k_{2}+k_{3}. $
Note that  terms in (\ref{bispectrum1}) contain slow-roll parameters (\ref{eq:slowrollparameter}) and the first term in (\ref{eq:basicformula2}) is therefore suppressed by 
\begin{eqnarray}
	\epsilon \: (1-n_{s})
	=
	(2.1 \times 10^{-3} ) 
	\times  
	\left ( 
		\frac{\epsilon}{0.05}\right ) \left ( \frac{1-n_{s}}{0.042}
	\right ),
\end{eqnarray}
and also by a similar factor of  $\eta (1-n_{s})$.

The two-scalar-one-graviton correlation was also worked out in Ref. \cite{maldacena1} as 
\begin{eqnarray}
	\langle   
		\zeta _{\vec{k}_{1}}  \zeta_{\vec{k}_{2}}  \gamma_{\vec q}^{s} 
	\rangle'
	&=&
	\frac{1}{4 \epsilon }  
	\left ( \frac{H_{*} }{M_{\rm pl} } \right )^{4}
	\frac{1}{q^{3} k_{1}^{3}k_{2}^{3}} 
	\varepsilon_{ij}^{s} (\vec{q}\,) 
	{(\vec{k}_{1})}^{i}  (\vec{k}_{2}) ^{j}
	\left (
		-L 
		+
		\frac{qk_{1} + k_{1}k_{2} + k_{2}q }{L} 
		+
		\frac{qk_{1}k_{2} }{L^{2}}
	\right ), 
	\nonumber
\end{eqnarray}
where $L=k_{1}+k_{2}+q$.   The third term in (\ref{eq:basicformula2}) thus turns out to  be a product of
\begin{eqnarray}
	\frac{
		\langle   
			\zeta _{\vec{k}_{1}}  \zeta_{\vec{k}_{2}}  \gamma_{\vec q}^{s} 
		\rangle'
	}{P_{\zeta} (k_{1}) P_{\zeta}(k_{2})  }
	&=&
	4 \epsilon\, \varepsilon_{ij}^{s} (\vec{q}\, ) 
	{(\vec{k}_{1})}^{i} (\vec{k}_{2}) ^{j}
	\frac{1}{q^{3}}
	\left (
		-L 
		+
		\frac{qk_{1} + k_{1}k_{2} + k_{2}q }{L} 
		+
		\frac{qk_{1}k_{2} }{L^{2}}
	\right )
	\nonumber
\end{eqnarray}
and 
\begin{eqnarray}
	\frac{
		\langle   
			\zeta _{\vec{k}_{3}}  \zeta_{\vec{k}_{4}}  \gamma_{-\vec q}^{s} 
		\rangle'
	}{P_{\gamma } (q)}
	&=&
	P_{\zeta}(k_{3})  
	\varepsilon_{lm}^{s} (- \vec{q}\,)
	{(\vec{k}_{3})}^{l}  (\vec{k}_{4}) ^{m}
	\frac{1}{ 4 k_{4}^{3}} 
	\left (
		-M 
		+
		\frac{qk_{3} + k_{3}k_{4} + k_{4}q }{M} 
		+
		\frac{qk_{3}k_{4} }{M^{2}}
	\right ).
\end{eqnarray}
Here we have defined $M=k_{3}+k_{4}+q$. It is now clear that the  suppression factor in the third term in (\ref{eq:basicformula2}) is simply $\epsilon$. If $(1-n_{s})  \approx 0.042$, it may be possible that the third term is more dominant than the other two which are suppressed either by $\epsilon(1-n_{s})$,  $\eta (1-n_{s})$ or $(1-n_{s})^{2}$. Thus, the graviton exchange effect could be seen in the future observation of the trispectrum.


\end{document}